\documentclass[%
superscriptaddress,showkeys,
12pt,
amsmath,amssymb,
aps,
]{revtex4-2}

\usepackage{graphicx}
\usepackage{dcolumn}
\usepackage{bm}
\usepackage{xcolor}
\usepackage{url}
\usepackage{makecell}
\usepackage{subfigure}
\usepackage{multirow}
\usepackage{booktabs}
\usepackage{verbatim}
\usepackage{diagbox}
\usepackage[marginal]{footmisc}
\usepackage{amssymb}

\begin{document}


\title{ Exploring the Impact of anti-shadowing effect on Unintegrated Gluon Distributions in the MD-BFKL Equation} 
\author{Xiaopeng Wang}
\affiliation{Institute of Modern Physics, Chinese Academy of Sciences, Lanzhou 730000, China}
\affiliation{School of Nuclear Science and Technology, Lanzhou University, Lanzhou 730000, China}
\affiliation{School of Nuclear Science and Technology, University of Chinese Academy of Sciences, Beijing 100049, China}

\author{Yanbing Cai}
\email{yanbingcai@mail.gufe.edu.cn (corresponding author)}
\affiliation{Guizhou Key Laboratory in Physics and Related Areas, Guizhou University of Finance and Economics, Guiyang 550025, China}
 \affiliation{Southern Center for Nuclear-Science Theory (SCNT), Institute of Modern Physics, Chinese Academy of Sciences, Huizhou 516000, China}

\author{Xurong Chen}
\email{xchen@impcas.ac.cn (corresponding author)}
\affiliation{Institute of Modern Physics, Chinese Academy of Sciences, Lanzhou 730000, China}
\affiliation{School of Nuclear Science and Technology, University of Chinese Academy of Sciences, Beijing 100049, China}
\affiliation{Southern Center for Nuclear-Science Theory (SCNT), Institute of Modern Physics, Chinese Academy of Sciences, Huizhou 516000, China}
\date{\today}

\begin{abstract}
This paper presents a comprehensive analysis of the MD-BFKL equation, considering both shadowing and anti-shadowing effects in gluon recombination processes. By deriving analytical expressions for unintegrated gluon distributions through the solution of the MD-BFKL equation, with and without the incorporation of anti-shadowing effect, we offer new insights into the influence of these effects on the behavior of unintegrated gluon distributions. Our results, when compared to the CT18NLO gluon distribution function, demonstrate that the anti-shadowing effect has a notably stronger impact on the characteristics of unintegrated gluon distributions, particularly in regions of high rapidity and momentum. This work significantly contributes to the understanding of gluon recombination mechanisms and their implications in high energy physics.
\end{abstract}
\maketitle


\section{introduction}
In the framework of perturbative quantum chromodynamics (pQCD), the understanding and prediction of parton distribution functions (PDFs) in protons can be achieved through the utilization of QCD evolution equations. The well-established Dokshitzer-Gribov-Lipatov-Altarelli-Parisi (DGLAP) evolution equation~\cite{Dokshitzer:1977sg,Altarelli:1977zs,Gribov:1971zn} is applied to describe the growth of partons at high $Q^2$, while the Balitsky-Fadin-Kuraev-Lipatov (BFKL) equation~\cite{Lipatov:1976zz,Fadin:1975cb,Kuraev:1976ge,Kuraev:1977fs,Balitsky:1978ic,Balitsky:1979ns} provides insights into gluon splitting in the small-$x$ region~\cite{Mukherjee:2023snp}. 

Nonetheless, both the DGLAP and BFKL equations exhibit unbounded growth of the gluon density, which violates either the unitary or the Froissart bound~\cite{Froissart:1961ux}. To address this issue, higher-order corrections must be taken into consideration to regulate this uncontrolled growth. Specifically, in regions of high density, gluons recombine, and such recombination effects become particularly significant in the near-unitarity limit, eventually leading to saturation. The Color Glass Condensate (CGC) effective theory serves as a potent and valuable tool for describing the saturation phenomenon~\cite{Iancu:2000hn,Iancu:2001ad,Jalilian-Marian:2005ccm,Gelis:2010nm}. Within the framework of CGC, the Jalilian-Marian-Iancu-McLerran-Weigert-Leonidov-Kovner (JIMWLK) renormalization group equation~\cite{Jalilian-Marian:1997qno,Iancu:2000hn,Weigert:2000gi,Mueller:2001uk} is introduced to account for small-$x$ QCD evolution incorporating non-linear corrections. The mean-field approximation of JIMWLK, known as the Balitsky-Kovchegov (BK) equation~\cite{Balitsky:1995ub,Kovchegov:1999yj,Cai:2023iza}, serves as a non-linear extension of the BFKL equation. By considering the dipole scattering amplitude, the BK equation bridges the gap between the unsaturated and saturated regions.

Furthermore, an alternative approach for modifying the gluon DGLAP evolution equation has been proposed by Gribov-Levin-Ryskin and Mueller-Qiu (GLR-MQ) equation ~\cite{Gribov:1983ivg,Mueller:1985wy,Devee:2014fna,Cai:2024guq,Boroun:2023clh,Boroun:2023zvc}. However, it is worth noting that the derivation of the GLR-MQ equation relies on the application of AGK-cutting rules~\cite{Abramovsky:1973fm}. As highlighted in Ref.~\cite{Zhu:1998hg}, the employment of AGK-cutting rules in the GLR-MQ equation presents certain limitations and drawbacks. Notably, the GLR-MQ equation overlooks the antishadowing corrections associated with gluon recombination. To address these limitations, {Zhu, Ruan and Shen} revisited parton recombination within the framework of the QCD evolution equation using time-ordered perturbation theory {(TOPT)}. This led to the derivation of the modified BFKL equation (MD-BFKL) and the modified DGLAP equation (MD-DGLAP)~\cite{Zhu:1999ht,Ruan:2006ud}. Importantly, both the antishadowing effects and shadowing effects are considered within these modified equations, ensuring the preservation of momentum conservation throughout the QCD recombination processes~\cite{Zhu:1993rr,Zhu:1994uw,Zhu:1995np,Zhu:1996nq}.

In Ref.~\cite{Ruan:2006ud},  the numerical solution of MD-BFKL demonstrated that anti-shadowing effect has a sizable impact on the gluon distribution. Therefore, it is meaningful to get an analytic unintegrated gluon   distribution from MD-BFKL. In recent years, a solution of kinematic-constraint improved MD-BFKL equation (KC-MD-BFKL) is obtained~\cite{Phukan:2018gnk}  in the near saturation region. In fact, based on variable transformation, MD-BFKL equation can be transformed into BK-like equation \cite{Ruan:2006ud}. The BFKL kernel in BK-like equation can be expanded to a differential kernel as done for the BK equation \cite{Munier:2003vc,Peschanski:2005wg,Marquet:2005ic}, which is connected to the Fisher-Kolmogorov-Petrovsky-Piscounov (FKPP) equation~\cite{fkpp,LAU198544,Bacaër2011}. Therefore, the MD-BFKL equation can be rewritten as an analytically solvable nonlinear partial differential equation under the diffusive approximation. In this paper, we shall solve this nonlinear partial differential equation analytically with and without the inclusion of the anti-shadowing effect. In our following calculations, it is shown that the anti-shadowing correction plays an important role for unintegrated gluon distribution.

The arrangement of this article is as follows. In Sec.~\ref{sec:BK-like}, we briefly introduce the MD-BFKL equation and its transformation form, the BK-like equation. In Sec.~\ref{sec:solution_bk_like equation}, the analytic solutions of BK-like equation with and without the anti-shadowing effect are obtained.  In Sec.~\ref{sec:match_gluon}, the results of fitting to gluon distribution function by using our solutions are provided. In Sec.~\ref{sec:conclusion}, the discussion and summary are given.

\section{BK-like equation from modified BFKL equation}
\label{sec:BK-like}
In Ref.\cite{Ruan:2006ud}, a unitarized MD-BFKL equation incorporating both shadowing and anti-shadowing corrections of the gluon recombination is shown as
\begin{small}
	\begin{equation}
		\label{eq:md-bfkl}
		\begin{aligned}-x \frac{\partial F\left(x, \textbf{k}^2\right)}{\partial x}= & \frac{\alpha_s N_c \textbf{k}^2}{\pi} \int_{\textbf{k}_{\min }^{\prime 2}}^{\infty} \frac{d \textbf{k}^{\prime 2}}{\textbf{k}^{\prime 2}}\left\{\frac{F\left(x, \textbf{k}^{\prime 2}\right)-F\left(x, \textbf{k}^2\right)}{\left|\textbf{k}^{\prime 2}-\textbf{k}^2\right|}+\frac{F\left(x, \textbf{k}^2\right)}{\sqrt{\textbf{k}^4+4 \textbf{k}^{\prime 4}}}\right\}\\
			&-\frac{36 \alpha_s^2}{\pi \textbf{k}^2 R^2} \frac{N_c^2}{N_c^2-1} F^2\left(x, \textbf{k}^2\right)+\frac{18 \alpha_s^2}{\pi \textbf{k}^2 R^2} \frac{N_c^2}{N_c^2-1} F^2\left(\frac{x}{2}, \textbf{k}^2\right),\end{aligned}
	\end{equation}
\end{small}
where $F\left(x, \textbf{k}^2\right)$ is unintegrated gluon distribution~\cite{Boroun:2023ldq,Boroun:2023goy,Kutak:2004ym,Bolognino:2018mlw}, $\alpha_s$ is the coupling constant, $N_c$ is the color number and $R$ is the effective correlation length of two recombination gluons. $R\approx 5 \,\mathrm{GeV}^{-1}$ stands for the hadronic characteristic radius, while $R\approx 2 \,\mathrm{GeV}^{-1}$ stands for the hot spot radius~\cite{Askew:1992tw}. In this work, $R$ is determined by fitting gluon distribution function. The linear part in Eq.$\,$(\ref{eq:md-bfkl}) is the famous BFKL kernel. There are two nonlinear terms (second term and third term on the r.h.s) in Eq.$\,$(\ref{eq:md-bfkl}). The second term stands  for shadowing effect and the second term stands for anti-shadowing effect, respectively. The total nonlinear parts are the complete contribution of the gluon recombination.
{The MD-BFKL equation is a kind of integral-partial differential equation, which is difficult to solve analytically. Fortunately, a  transformation method has been proposed in~\cite{Ruan:2006ud} to reduce the MD-BFKL equation by defining}
\begin{equation}
	\label{eq:N-F}
	N(x,\textbf{k}^2)\equiv\frac{27\alpha_s}{4\textbf{k}^2R^2}F(x,\textbf{k}^2).
\end{equation}
With this definition, Eq.$\,$(\ref{eq:md-bfkl}) can be rewritten as
	\begin{equation}
	\label{eq:md-bk}
	\begin{aligned}
		-x \frac{\partial N\left(\textbf{k}^2, x\right)}{\partial x}&=
		 \frac{\alpha_s N_c}{\pi} \int_{\textbf{k}_{\min }^{\prime 2}}^{\infty} \frac{d \textbf{k}^{\prime 2}}{\textbf{k}^{\prime 2}}
		 \left\{\frac{\textbf{k}^{\prime 2} N\left(\textbf{k}^{\prime 2}, x\right)-\textbf{k}^2 N\left(\textbf{k}^2, x\right)}{\left|\textbf{k}^{\prime 2}-\textbf{k}^2\right|}+\frac{\textbf{k}^2 N\left(\textbf{k}^2, x\right)}{\sqrt{\textbf{k}^4+4 \textbf{k}^{\prime 4}}}\right\} \\
		&-2 \frac{\alpha_s N_c}{\pi} N^2\left(\textbf{k}^2, x\right)+\frac{\alpha_s N_c}{\pi} N^2\left(\textbf{k}^2, \frac{x}{2}\right) .
	\end{aligned}
\end{equation}
{Here we refer to Eq.$\,$(\ref{eq:md-bk}) as the BK-like equation, which exhibits a comparable structure to the conventional BK equation in momentum space. Therefore, $N$ can be called ``dipole scattering amplitude".}

Now Eq.$\,$(\ref{eq:md-bk}) can be simplified into an analytically solvable form. It is well-known that the BFKL kernel in Eq.$\,$(\ref{eq:md-bk}) can be transformed into a differential kernel \cite{Kovchegov:1999yj}. Therefore, Eq.$\,$(\ref{eq:md-bk}) can be written as 
\begin{equation}
	\label{eq:diff_md-bk}
	\frac{1}{\bar{\alpha}}\frac{\partial N({L},{Y})}{\partial {Y}}=\chi(-\partial_{L})N({L},{Y})-2N^2({L},{Y})+N^2({L},{Y}+\mathrm{ln}2),
\end{equation}
where {$\bar{\alpha}=\frac{\alpha_sN_c}\pi$}, ${Y}=\mathrm{ln}\frac{1}{x}$, ${L}=\mathrm{ln}(\frac{k^2}{k^2_0})$, $k^2_0$ is an arbitrary scale and $\chi(-\partial_{L})$ is the BFKL kernel \cite{Lipatov:1976zz,Balitsky:1978ic,Kuraev:1977fs} ,
\begin{equation}
\chi\left(-\partial_{L}\right)=2\psi(1)-\psi\left(-\partial_{L}\right)-\psi\left(1+\partial_{L}\right).
\end{equation}
{$\chi\left(-\partial_{L}\right)$ can be expand around critical point $\gamma_c$~\cite{Marquet:2005ic},
\begin{equation}
\label{eq:expands}
\begin{gathered}
		\chi(-\partial_{L}) =\sum_{p=0}^P\frac{\chi^{(p)}(\gamma_c)}{p!}(-\partial_L-\gamma_c)^p \\
		=\sum_{p=0}^P(-1)^pA_p\partial_L^p. 
\end{gathered}
\end{equation}}
Expanding $\chi(-\partial_{L})$ to the second order (the diffusive approximation), Eq.$\,$(\ref{eq:diff_md-bk}) is rewritten as~\cite{Munier:2003vc,Marquet:2005ic,Peschanski:2005wg}
\begin{equation}
	\label{eq:2_order_diff_md-bk}
	A_0{N}-2N^2+{N^2({L},{Y}+\mathrm{ln}2)}-\frac{1}{\bar{\alpha}_s} \frac{\partial{N}}{\partial {Y}}-A_1\frac{\partial{N}}{\partial {L}}+A_2\frac{\partial^2{N}}{\partial^2{L}}=0.
\end{equation}
Here $A_0$, $A_1$ and $A_2$ are  the expansion  coefficients. For the sake of simplicity, we do not write the default variables $L$ and $Y$ for $N$. In the next section, we shall simplify ${N^2(L,Y+\mathrm{ln}2)}$ in Eq.$\,($\ref{eq:2_order_diff_md-bk}) to get the analytical solution.

\section{ANALYTICAL SOLUTION OF BK-like EQUATION}
\label{sec:solution_bk_like equation}
In this section, the detailed description of solving Eq.$\,($\ref{eq:2_order_diff_md-bk}) with fixed coupling is provided. In order to consider the impact of anti-shadowing correction, we shall take into account the two strategies for the nonlinear parts.

(I). The anti-shadowing term is canceled directly. Only shadowing effect is considered. Therefore, Eq.~$\,($\ref{eq:2_order_diff_md-bk}) is rewritten as
	\begin{equation}
		\label{eq:2_order_diff_md-bk_only_sh}
		A_0{N}-2N^2-\frac{1}{\bar{\alpha}_s} \frac{\partial{N}}{\partial  Y}-A_1\frac{\partial{N}}{\partial {L}}+A_2\frac{\partial^2{N}}{\partial^2{L}}=0.
	\end{equation}
%
By the variable substitution,
\begin{equation}
	\label{eq:t_Y_x_L}
	\begin{aligned}
		t&=A_0\bar{\alpha}_s{Y},\\
		x&=\sqrt{\frac{A_0}{A_2}}({L}-A_1\bar{\alpha}_s{Y}),\\
			{N}&=A_0\bar{u}(x,t).
   \end{aligned}
\end{equation}

Eq.$\,$(\ref{eq:2_order_diff_md-bk_only_sh}) is transformed into FKPP equation,
\begin{equation}
	\label{eq:fkpp}	
	\bar{u}_{xx}-\bar{u}_t+\bar{u}-2\bar{u}^2=0.
\end{equation}
Eq.~(\ref{eq:fkpp}) means that $\frac{F(x.k^2)}{k^2}$ satisfied reaction-diffusion FKPP equation when only shadowing term is retained.
It is easy to get its solution with homogeneous balance method \cite{Wang:2020stj},
\begin{equation}
	\label{eq:fkpp_solution}
	\bar{u}=\frac{1}{2 \left(e^{2 \theta -\frac{5 t}{6}+\frac{x}{\sqrt{6}}}+1\right)^2}.
\end{equation}
where $\theta$ is arbitrary constant. Naturally, unintegrated gluon is obtained
\begin{equation}
	\label{eq:ugd_1}
	{F}({L},{Y})=\frac{4k_0^2 R^2}{27\alpha_s}e^{L} \left[\frac{A_0}{2\left(1+e^{2\theta +\sqrt{\frac{A_0}{6A_2}}{L}-\left(\sqrt{\frac{A_0}{6A_2}}A_1+\frac{5}{6}A_0 \right)\alpha_s {Y} } \right)^2}\right].
\end{equation}

(II). Both shadowing and anti-shadowing effects are retained. {In this situation, the only difficulty for getting analytic solutions is how to deal with $N^2({L},{Y}+\mathrm{ln}2)$.}
When we consider that Eq.$\,($\ref{eq:2_order_diff_md-bk}) has a traveling wave solution, the equation only depends on a single variable $\tau={L}+b{Y}$, where $b$ is an arbitrary constant. Therefore, ${N^2(\mathrm{L},Y+\mathrm{ln}2)} $ is written as ${N^2(\tau+{b}\cdot \mathrm{ln}2)}$.  Then expand ${N^2(\tau+{b}\cdot \mathrm{ln}2)}$ in a Taylor series around $\tau$ and retain up to the first order,
 \begin{equation}
 	\label{eq:approx_2_2}
 	N^2(\tau+b\cdot{\ln}2)\approx  N^2(\tau) +  b \cdot {\ln}2  \frac{\partial N^2(\tau)}{\partial \tau} . 
 \end{equation}

{Using variable $\tau$, Eq.$\,$(\ref{eq:2_order_diff_md-bk}) is rewritten as}
\begin{equation}
	\label{eq:approx_2_bk_origin}
	A_0 {N}-2{N}^2+N^2(\tau+b\ln2)-\frac{{b}}{\bar{\alpha}_{\mathrm{s}}} \frac{\partial {N}}{\partial \tau}-{A}_1 \frac{\partial {N}}{\partial \tau}+{A}_2 \frac{\partial^2 {N}}{\partial^2 \tau}=0.
\end{equation}
Then, when $N^2(\tau+b\ln2)$ is replaced by Eq.~(\ref{eq:approx_2_2}), Eq.~(\ref{eq:approx_2_bk_origin}) is approximated  as 
\begin{equation}
	\label{eq:approx_2_bk}
	A_0 {N}-{N}^2+2 b \ln 2 {N} \frac{\partial {N}(\tau)}{\partial \tau}-\frac{{b}}{\bar{\alpha}_{\mathrm{s}}} \frac{\partial {N}}{\partial \tau}-{A}_1 \frac{\partial {N}}{\partial \tau}+{A}_2 \frac{\partial^2 {N}}{\partial^2 \tau}=0.
\end{equation}

Eq.(\ref{eq:approx_2_bk}) can be solve by using the homogeneous balance method~\cite{FAN1998403}. 
{One can suppose $N(\tau)$ can be expanded to}
\begin{equation}
\label{eq:expand_N}
	N(\tau)=\sum_{i=0}^ma_iv^i(\tau),
\end{equation}
{where $a_i$ is the expansion coefficient and $v(\tau)$ is the expansion function. The condition for $v(\tau)$ is $\frac{\mathrm{d}v}{\mathrm{d}\tau}=k\left(1-v^2\right) $. Using the homogeneous balance method to balance the highest power of $v$ from $\frac{\partial^2 {N}}{\partial^2 \tau}$ with the highest power of $v$ from the nonlinear terms $N^2$, We can get
\begin{equation}
	m+2=2m.
\end{equation}}
{It is obvious that $m=2$.
Therefore, the form of solution of Eq.$\,$(\ref{eq:approx_2_bk}) is,}

\begin{equation}
	\label{eq:assume_solution_apprax_2}
	\begin{aligned}
		& N(\tau)=a_0+a_1 v(\tau)+ a_2 v(\tau)^2, \\
		& v(\tau)^{\prime}=k\left(1-v(\tau)^2\right),\\
		& v(\tau)= \mathrm{tan}(k\tau),
	\end{aligned}
\end{equation}
where $a_0$, $a_1$, $a_2$ and $k$ are parameters to be determined. Then, substitute Eq.~(\ref{eq:assume_solution_apprax_2}) into Eq.~(\ref{eq:approx_2_bk}) and make the coefficients of $v^i$ zero,
\begin{small}
	\begin{equation}
		\label{eq:algebraic equ}
		\begin{aligned}
			&-a^2_0+a_0A_0-a_1A_1k-\frac{a_1bk}{\alpha_s}+2a_2A_2k^2+2a_0a_1bk\mathrm{ln}2=0,\\
			&-2a_0a_1+A_0a_1-2A_1a_2k-\frac{2a_2bk}{\alpha_s}-2a_1A_2k^2+2a_1^2bk\mathrm{ln}2+
			4a_0a_2bk\mathrm{ln}2=0,\\			
			&-a_1^2-2a_0a_2+A_0a_2+a_1A_1k+\frac{a_1bk}{\alpha_s}-8a_2A_2k^2-2a_0a_1bk\mathrm{ln}2+
			6a_1a_2bk\mathrm{ln}2=0,\\
			&-2a_1a_2+2A_1a_2k+\frac{2a_2bk}{\alpha_s}+2a_1A_2k^2-2a_1^2bk\mathrm{ln}2-4a_0a_2bk\mathrm{ln}2+
			4a_2^2bk\mathrm{ln}2=0,\\
			&-a_2^2+6a_2A_2k^2-6a_1a_2bk\mathrm{ln}2=0,\\
			&a_2=0.
		\end{aligned}
	\end{equation}
\end{small}
By solving Eq.~($\ref{eq:algebraic equ}$), we get a suitable set of parameter values,
\begin{equation}
	\label{eq:sulution_algebraic}
	\begin{aligned}
		a_0&=\frac{A0}{2},\\
		a_1&=\frac{A0}{2},\\
		a_2&= 0,\\
		k&=\frac{b(\mathrm{ln}2)A_0}{2A_2},\\
		A_1&=\frac{a_1\alpha_s-bk+A_0\alpha_sbk\mathrm{ln}2}{\alpha_s k}.
	\end{aligned}
\end{equation}
Therefore, the solution of Eq.~(\ref{eq:approx_2_bk}) is obtained,
\begin{equation}
	\label{eq:solution_approx2_bk}
	{N}=\frac{A_0}2\text{Tanh}(\frac{{b}(\ln2){A}_0}{2{A}_2}\tau+\theta)+\frac{{A}_0}2,
\end{equation}
where $\theta$ is a free parameter. Using the definition in Eq.~(\ref{eq:N-F}), the unintegrated gluon distribution is written as
\begin{equation}
	\label{eq:ugd_2}
	F({L},{Y})=\frac{4k_0^2 R^2}{27\alpha_s}e^{L}\left[\frac{A_0}2\text{Tanh} \left(\frac{{b}(\ln2){A}_0}{2{A}_2}(L+bY)+\theta \right)+\frac{{A}_0}2\right].
\end{equation}


Generally, the values of $A_0$, $A_1$ and $A_2$ are determined by the saddle point in the diffusive approximation, however, to enhance the description of the gluon distribution, we have relaxed this constraint. Therefore, in next section, the parameters in Eq.~(\ref{eq:ugd_1}) and Eq.~(\ref{eq:ugd_2}) are determined by fitting to the gluon distribution from the database. The gluon distribution exhibits variability across different databases. For this study, CT18NLO was selected because it considers the saturation effect in small-$x$ by introducing the $x$ dependent scale.  
\begin{table}[htbp]
	\caption{\label{tab:table1}%
		Parameters from the fit to the CT18NLO~\cite{Guzzi:2021fre,Buckley:2014ana} with and without anti-shadowing.
	}
		\begin{ruledtabular}
		\begin{tabular}{ccc}
			&shadowing and anti-shadowing&shadowing \\
			\colrule
			\textrm{$A_0$} &0.517 $\pm \,0.010$&	1.507 $\pm\,0.020$\\	
			\textrm{$A_1$}& - & 0.308 $\pm\,0.023$\\
			\multicolumn{1}{c}{$A_2$}&0.137 $\pm \,0.002$& 0.541 $\pm\,0.007$\\
			\textrm{$b$}&-0.431 $\pm \,0.003$&-\\
			\textrm{$R$}&2.432 $\pm \,0.030$&2.367 $\pm\,0.029$\\
			$\chi^2/dof$&$37.660/(1260-5)$&$36.420/(1260-5)$\\
		\end{tabular}
		\end{ruledtabular}
\end{table}
\begin{figure}[hpbt]
	\includegraphics[scale=1.2]{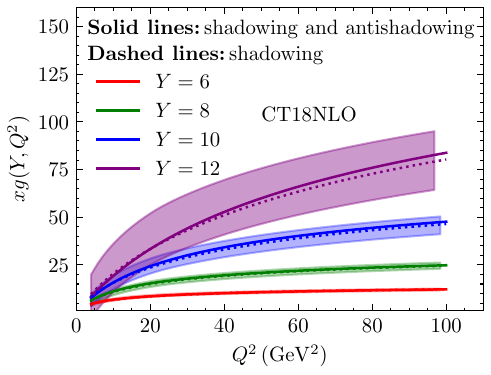}
	\caption{\label{fig:xg} 
    The gluon distribution function fitted by CT18NLO~\cite{Guzzi:2021fre,Buckley:2014ana} using Eq.~(\ref{eq:xg_ugd}) at $ Y=6$, $ Y=8$, $ Y=10$ and $ Y=12$ {(from the bottom to top).} Color bands stand for the CT18NLO gluon data with uncertanties. The solid lines represent that both shadowing effect and anti-shadowing effect are considered. The dashed lines indicate that only shadowing effect is included. }
\end{figure}

\begin{figure}[hpbt]
	\includegraphics[scale=1.2]{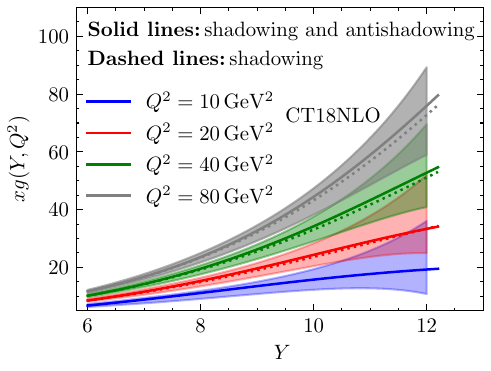}
	\caption{\label{fig:xg_Y}  The gluon distribution function fitted by CT18NLO~\cite{Guzzi:2021fre,Buckley:2014ana} using  Eq.~(\ref{eq:xg_ugd}) at $ Q^2=10\,\mathrm{GeV^2}$, $ Q^2=20\,\mathrm{GeV^2}$, $ Q^2=40\,\mathrm{GeV^2}$ and $ Q^2=80\,\mathrm{GeV^2}$ {(from the bottom to top).} Color bands stand for the CT18NLO gluon data with uncertanties. The solid lines represent that both shadowing effect and anti-shadowing effect are considered. The dashed lines indicate that only shadowing effect is included. }
\end{figure}

\begin{figure}[hpbt]
	\includegraphics[scale=1.2]{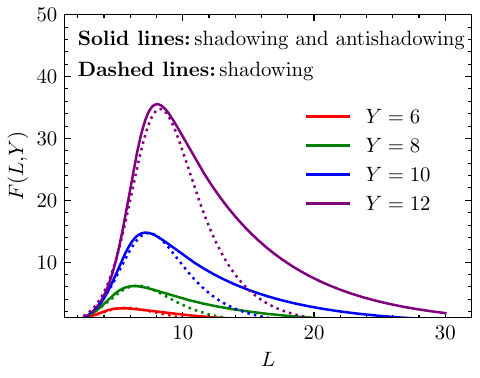}
	\caption{\label{fig:ugd} Comparison of the unintegrated gluon distribution function including shadowing and anti-shadowing effects with the unintegrated gluon distribution function only including the shadowing effect as a function of $ Y$  at $ Y=6$, $ Y=8$, $ Y=10$ and $ Y=12$ {(from the bottom to top).} The solid lines represent that both shadowing effect and anti-shadowing effect are considered. The dashed lines indicate that only shadowing effect is included.}
\end{figure}

\begin{figure}[hpbt]
	\includegraphics[scale=1.2]{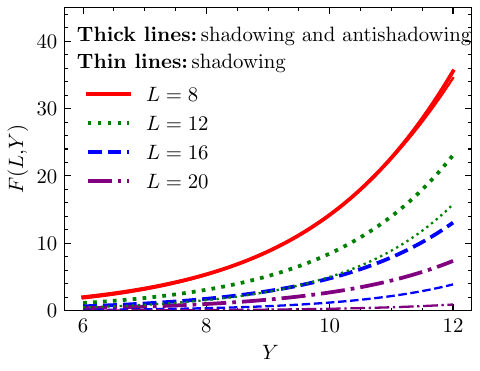}
	\caption{\label{fig:ugd_Y} Comparison of the unintegrated gluon distribution function including shadowing and anti-shadowing effects with the unintegrated gluon distribution function only including the shadowing effect as a function of $ Y$ at $ L=8$, $ L=12$, $ L=16$ and $ L=20$ respectively. {Thick lines represents the situation where both shadowing and anti-shadowing effects are considered. Thin lines stand for the situation where only the shadowing effect is included.}}
\end{figure}

\begin{figure}[hpbt]
	\includegraphics[scale=1.2]{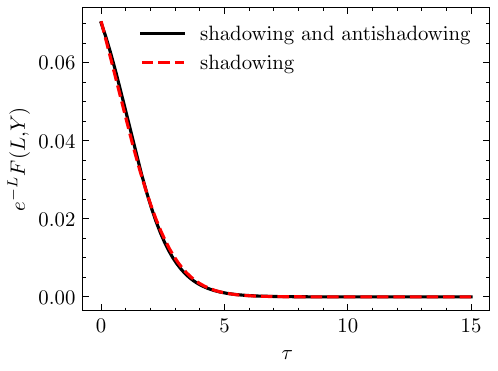}
	\caption{\label{fig:elugd} $ e^{-{L}}F({L},{Y})$ as a function of only one dimensionless variable $\tau$. The solid line represents that both shadowing effect and anti-shadowing effect are considered. The dashed line indicates that only shadowing effect is included.}
\end{figure}

\section{the gluon distribution functions }
\label{sec:match_gluon}
The relation of the unintegrated gluon distribution $F({Y},{L})$ with the integrated gluon distribution ${xg}({Y},{Q^2})$ is
\begin{equation}
	\label{eq:xg_ugd}
	{xg}({Y},{Q^2})=\int_{0}^{{L'}=\frac{{Q^2}}{k^2_0}}F({Y},{L}){\mathrm{d}L}.
\end{equation}
In order to get a definite solution which can match gluon distribution function, the free parameters in Eq.~(\ref{eq:ugd_1}) and Eq.~(\ref{eq:ugd_2}) are obtained by fitting CT18NLO~\cite{Guzzi:2021fre,Buckley:2014ana}. As we are focused on the properties of gluon distribution in the small-$x$ region, the fitting range of $Y$ is set from 6 to 12. The fitting range of $Q^2$ is from 4~$\mathrm{GeV^2}$ to 100~$\mathrm{GeV^2}$.  We set $\alpha_s=0.2$ and $k_0^2=0.04\, \mathrm{GeV^2}$. The coefficients from fitting to CT18NLO are shown in Table~\ref{tab:table1}. Parameter values that are not presented in the Table~\ref{tab:table1} can be obtained by the relationship in Eq.~(\ref{eq:sulution_algebraic}). 

 As shown in Fig.~\ref{fig:xg}, the fitting results of two situations for integrated gluon distribution $ xg(Y,Q^2)$ are presented at $ Y=6$, $ Y=8$, $ Y=10$ and $ Y=12$ respectively. And Fig.~\ref{fig:xg_Y} demonstrates $ xg(Y,Q^2)$ as a function of $ Y$ at $ Q^2=10\,\mathrm{GeV^2}$, $Q^2=20\,\mathrm{GeV^2}$, $ Q^2=40\,\mathrm{GeV^2}$ and $ Q^2=80\,\mathrm{GeV^2}$ respectively. Our solutions can be related to the gluon distribution by using Eq.~(\ref{eq:xg_ugd}). The solid lines represent that both shadowing effect and anti-shadowing effect are considered. The dashed lines indicate that only shadowing effect is included. Despite fitting the same gluon data, we can still find that the differences between the gluon distribution functions with and without anti-shadowing effect as $Q^2$ increases, especially in the higher $ Y$ region.{ When $Y$ is larger, total effect from shadowing and anti-shadowing terms is more obvious, which means that recombination effect becomes more and more important when gluons approach or enter the saturation region. As gluons approach or enter the deep saturation region, the dominant effect is the shadowing resulting from the recombination process, which suppresses the growth of gluons to maintain unitarity.  When gluons move away from the deep saturation region, the contribution of anti-shadowing effect from recombination process is visible. the shadowing effects is weakened by the anti-shadowing effects. Therefore, the value of gluons distribution with anti-shadowing effects is higher than that without anti-shadowing effects as $Q^2$ increases.}
 
The unintegrated gluon distributions $F({Y,L})$ with and without anti-shadowing effect are shown in Fig.~\ref{fig:ugd} and Fig.~\ref{fig:ugd_Y}. Fig.~\ref{fig:ugd} illustrates $F({Y,L})$ as a function of $ L$ at $ Y=6$, $ Y=8$, $ Y=10$ and $ Y=12$ respectively. Figure~\ref{fig:ugd_Y} demonstrates the behavior of $F({Y,L})$ with $ Y$ at $ L=8$, $ L=12$, $ L=16$ and $ L=20$ respectively. For the evolution of unintegrated gluon distributions, the impact of anti-shadowing effect is significant. In large $ L$, the unintegrated gluon distributions with anti-shadowing effect decreases more slowly. {When the gluon moves away from the saturation, the anti-shadowing effect gradually increases effect as $L$ becomes larger. } Therefore, the anti-shadowing correction is important for the gluon distribution.

It should note that, ${e^{-L}}F({L},{Y})\propto N({L},{Y})$ from Eq.~(\ref{eq:N-F}) has scaling characteristic when running coupling constant is fixed. In other word, ${e^{-L}}F({L},{Y})$ is a function of only one dimensionless variable $\tau={L}+b{Y}$. For the situation where only shadowing effect is considered, $\tau={L}+b_{sh}{Y}$ , where $b_{sh}=-(A_1+5\sqrt{\frac{A_0A_2}{6}})\alpha_s=- 0.4302$ which is close to the value in the case with anti-shadowing (see Table~\ref{tab:table1}). In general, as shown in Fig.~\ref{fig:elugd}, the difference for scaling behavior of two solutions is not obvious.{ Actually, the scaling property of ${e^{-L}}F({L},{Y})$ implies that ${e^{-\ln \frac{Q^2}{\Lambda^2}}}xg(x,Q^2)$ have geometric scaling characteristic when $\alpha_s$ is fixed~\cite{Kwiecinski:2002ep,Cai:2024guq}. However, the $\alpha_s$ can only be considered a constant when $Q^2$ is large. The correlation for geometric scaling properties between unintegrated gluon distribution and integrated gluon distribution is obvious in the differential relationship
\begin{equation}
	\begin{aligned}
			e^{-L}F(Y,L)&=\left( \frac{\partial e^{-\ln\frac{Q^2}{\Lambda^2}}xg(Y,Q^2)}{\partial \ln\frac{Q^2}{\Lambda^2}}+e^{-\ln\frac{Q^2}{\Lambda^2}}xg(Y,Q^2)\right)\mid_{Q^2=k^2},\\
			&=\left( \frac{\partial e^{-\ln\frac{Q^2}{\Lambda^2}}xg(Y,Q^2)}{\partial \tau}+e^{-\ln\frac{Q^2}{\Lambda^2}}xg(Y,Q^2)\right)\mid_{Q^2=k^2}.
	\end{aligned}
\end{equation} 
}
{The scaling behavior of $e^{-L}F(Y,L)$, comes from the geometric scaling of gluon distribution from CT18NLO. As shown in Fig.~\ref{fig:xg} and Fig.~\ref{fig:xg_Y}, the difference between two situations for  $ e^{-\ln\frac{Q^2}{\Lambda^2}}xg(Y,Q^2)$ is small. The values of parameter $b$ in these two situations are nearly identical, indicating that the discrepancy in scaling behavior between the two solutions is not significant.}

\section{Conclusion}
\label{sec:conclusion}

In this study, the MD-BFKL equation is transformed to the BK-like equation by using the relationship between ``dipole scattering amplitude" and unintegrated gluon distribution. The BK-like equation arises from the expansion of the BFKL kernel to second order using the diffusive approximation, resulting in a nonlinear partial differential equation. By considering the nonlinear term, $N(\frac{x}{2},k^2)$, in two strategies (with and without anti-shadowing), we derive solutions for the MD-BFKL equation using the homogeneous balance method.

To investigate the impact of the anti-shadowing effect, we compare and contrast the obtained solutions, successfully determining their parameters by fitting them to the CT18NLO gluon distribution function in the small-$x$ region. Our findings, illustrated in Fig.~\ref{fig:xg} and Fig.~\ref{fig:xg_Y}, reveal that the gluon distribution without the anti-shadowing effect exhibits a gentler growth as $Q^2$ and $Y$ increase. Fig.~\ref{fig:ugd} and Fig.~\ref{fig:ugd_Y} further demonstrate a significant disparity between the unintegrated gluon distributions with and without the anti-shadowing effect, with the latter showing a slower decrease in large $L$ values. This disparity becomes more pronounced with an increase in $Y$.

{In addition, the scalling behavior of ${e^{-L}}F({L},{Y})$ is presented. In fact, the scaling properties of ${e^{-L}}F({L},{Y})$ reflect geometric scalling property of $ e^{-\ln\frac{Q^2}{\Lambda^2}}xg(Y,Q^2)$ when $\alpha_s$ is fixed. }In conclusion, our obtained solutions contribute valuable insights into the nature of gluon distributions and pave the way for future phenomenological studies.


\begin{acknowledgments}
This work is supported by the Strategic Priority Research Program of Chinese Academy of Sciences under the Grant NO.~XDB34030301, Guangdong Major Project of Basic and Applied Basic Research NO.~2020B0301030008, Education Department of Guizhou Province under Grant No. QJJ[2022]016 and Guizhou Provincial Basic Research Program (Natural Science) under Grant NO.QKHJC-ZK[2023]YB027.

\end{acknowledgments}

\vfill
\newpage
\nocite{*}
\bibliographystyle{apsrev4-2}
\bibliography{apssamp}

\end{document}